\PassOptionsToPackage{table}{xcolor}
\RequirePackage{fix-cm}
\documentclass[sigconf]{acmart}                    % onecolumn (standard format)

\renewcommand\footnotetextcopyrightpermission[1]{} % removes conference venue

\usepackage{graphicx}
\usepackage{svg}
\usepackage{natbib}
\usepackage{longtable}
\usepackage{url}
\usepackage{booktabs}
\usepackage{enumitem}
\usepackage{tabularx}
\usepackage{wrapfig}
\usepackage[ruled,linesnumbered,nofillcomment]{algorithm2e}
\usepackage[table]{xcolor}
\usepackage{subcaption}
\usepackage{multicol}
\usepackage{multirow}
\usepackage{amsmath}
\usepackage{amsfonts}
\usepackage{bbm}
\usepackage{tcolorbox}
\usepackage{colortbl}

\newcolumntype{L}{>{\raggedright\arraybackslash}X}
\newtcolorbox{blockquote}{colback=gray!5!white,boxrule=0.4pt,colframe=gray!60!black,fonttitle=\bfseries}
\newcommand{\gray}[1]{\cellcolor{gray!20}\textbf{#1}}
\acmConference[MSR 2022]{MSR '22: Proceedings of the 19th International Conference on Mining Software Repositories}{May 23–24, 2022}{Pittsburgh, PA, USA}
\begin{document}

\title{\textit{Old but Gold:} Reconsidering the value of   \\feedforward learners for software analytics
}
%\subtitle{A case study on deep learning for predicting Bugzilla  issue close time }


\author{Rahul Yedida}
\email{ryedida@ncsu.edu}
\affiliation{North Carolina State University}

\author{Xueqi Yang}
\email{xyang37@ncsu.edu}
\affiliation{North Carolina State University}

\author{Tim Menzies}
\email{timm@ieee.org}
\affiliation{North Carolina State University}

%\authorrunning{Short form of author list} % if too long for running he

\date{Received: date / Accepted: date}

\begin{abstract}
There has been an increased interest in the use of deep learning approaches for software analytics tasks. State-of-the-art techniques leverage modern deep learning techniques such as LSTMs, yielding competitive performance, albeit at the price of longer training times.

Recently, \citet{galke2021forget} showed that at least for image recognition, a decades-old feedforward neural network can match the performance of modern deep learning techniques. This motivated us to try the same in the SE literature. Specifically, in this paper, we apply feedforward networks with some preprocessing to two analytics tasks: issue close time prediction, and vulnerability detection. We test the hypothesis laid by \citet{galke2021forget}, that feedforward networks suffice for many analytics tasks (which we call, the ``Old but Gold'' hypothesis) for these two tasks.
For three out of five datasets from these tasks, we achieve new high-water mark
results
(that out-perform the prior state-of-the-art results)
and for a fourth data set, Old but Gold performed as well
as the recent state of the art.   Furthermore, the old but gold  results were obtained orders of magnitude faster than prior work. For example, for issue close time, old but gold found good predictors   in 90 seconds (as opposed to the newer methods,  which took 6 hours to run).

Our results supports the ``Old but Gold'' hypothesis and leads  to the following recommendation: try simpler alternatives before   more complex methods. At the very least, this will produce 
a baseline result against which researchers can compare some other, 
supposedly more sophisticated, approach. And in the best case, they will obtain useful results that are as good as anything else, in a small fraction of the effort.

To  support open science,   all our scripts and data are available on-line at \url{https://github.com/fastidiouschipmunk/simple}.
\end{abstract}

% \keywords{deep learning \and software engineering \and issue close time \and feedforward networks \and hyper-parameter optimization}

\maketitle

\section{Introduction}
\label{sec:intro}

% Recall the wisdom from one of the Python Enhancement Proposals, PEP20: ``Simple is better than complex; complex is better than complicated"\footnote{\url{https://www.python.org/dev/peps/pep-0020/}}. While preferable in theory, to what extent does simplicity work in practice, and how often is it applied? Our investigation is based on these questions, leading to a more elementary approach to artificial intelligence as applied to software engineering. 

As modern infrastructure allows for cheaper processing, it has inevitably led to the exploration  of more complex modeling.
For example, many
    software engineering researchers
are now using  deep learning methods~\citep{gao2020checking, hoang2019deepjit, liu2019deep, zhou2019devign,zhou2019devign,10.1145/3238147.3240471,8955829}.

One problem with deep learning is that it can be very slow to run \cite{jiang2018linear,le2011optimization,martens2010deep}.  For example, for the case
study of this paper, we estimate
that we would need 6 years of CPU time.  Such long runtimes can complicate many aspects of the scientific process (e.g. initial investigations, subsequent attempts at reproduction).

% As recent examples, consider the GPT-3 language model ~\citep{brown2020language} (with 175 billion parameters) and the recent FixEfficientNet model ~\citep{touvron2020fixing} (with 480 million parameters) proposed for image classification. Borrowing from the deep learning literature, several researchers have applied modern deep learning technology to solve software engineering problems ~\citep{zhou2019devign, 10.1145/3238147.3240471, 8955829}. As a result, the learn models tend to be overly complex for the task, leading to skepticism among the software engineering community ~\citep{Majumder2018500TF}. As an example, 

% Deep learning is not without its critics. Gary Marcus \cite{marcus2018deep} cautions that deep learning may be over-hyped to the extent that it runs ``fresh risk for seriously dashed expectations" that could blinker AI researchers from trying new ideas as well as bringing another AI winter.
% Our reading of the   literature
% is that
% this caution is not shared by SE
% researchers since there   are very few   case studies
% comparing state-of-the-art deep learning against much simpler methods.

Accordingly,   this paper checks if anything simpler than deep learner can handle
SE tasks. Outside of SE there is some suggestion that deep learning researchers have rushed on too far and have overlooked
the benefits of simpler neural architectures. For example, \citet{galke2021forget} offer an
\textit{Old but Gold} hypothesis; i.e. that in their rush to try new algorithms, researchers have overlooked the advantages of more traditional approaches.
% For example,
% recently, \citet{yedida2021value} revisited the feedforward learners for defect prediction. Their approach used oversampling with hyper-parameter optimization to achieve state-of-the-art results and terminated in minutes. These results motivated us to explore the possibility of feedforward learners as a viable baseline in software analytics.
In their work,  \citet{galke2021forget}  showed that for image classification, simple, decades-old feedforward networks (described in \S\ref{sec:deeplearning}) can perform as well as modern deep learning techniques,
at some small fraction of the computational cost.

Since deep learning is widely used in software engineering,  it seems prudent to check for old but gold effects in SE applications. In this paper we explore two standard software analytics problems  using older-style neural networks as well as the latest state-of-the-art deep learning algorithms.

% in this paper we show 
% While exploring very slow deep learning algorithms, 
% \citet{menzies2018500+} found that very simple,  a baseline method against which they could evaluate their
% new deep learning results. For that baseline, they turned
% to a simple SVMs.
% But then they found that their simpler baseline
% could achieve better results than the deep learning methods of the state-of-the-art.
% baselines to achieve better results over 500 times faster. In that paper, they achieves a state-of-the-art (SOTA) results 2,700 times faster than prior work in issue lifetime prediction.
% The particular tasks explored here  will be (a)
% {\em predicting the close time for  issues in a code repository} and (b) \textit{vulnerability detection}.

% % In both case studies, we use simpler, feedforward networks with hyper-parameter optimization and some preprocessing to achieve state-of-the-art results. Notably, our results outperform these prior, complex works, while being cheaper to run. 

% Our description  paper is structured around these research questions:

%The first two questions test the hypothesis for two different domains. We will describe our approaches, show how these simpler techniques outperform the more complex ones, and perform the comparison using statistical tests. The last question is more open-ended and asks what the SE community can learn from our work. These lessons learned will generalize to domains beyond the ones studied in this paper.

The experiments of this paper show that simpler methods than prior work are better for some domains.
Specifically, a simple extension to a 1980s-style feedforward neural network, which we call ``SIMPLE'',
runs much faster than prior work (90 seconds versus 6 hours for issue lifetime prediction).  Since they run faster,
feedforward networks  are more amenable to automatic   tuning methods.
Such tuning requires multiple runs of a learner~\cite{Tantithamthavorn16,fu2016tuning,agrawal2018better,agrawal2019dodge}  and so the {\em faster} the learner,
the {\em more} we can tune it (which we do in this paper).  Hence SIMPLE's feedforward
networks out-perform the prior work in issue lifetime prediction since the latter is fundamentally hard to customize to the task at hand.

The rest of this paper is structured as follows.
\S \ref{sec:background} presents the necessary background and \S \ref{sec:tasks} discusses the SE task under consideration.
%\S \ref{sec:deeplearning} introduces deep learning; \S \ref{sec:dl4se} presents our literature review of deep learning in software engineering. 
\S \ref{sec:method} discusses our proposed approach. Then, in \S \ref{results}, we show our results. We discuss the threats to the validity of our study in \S \ref{sec:threats}. In \S \ref{sec:discussion} we conclude that  
before analysts  try very sophisticated
(but very slow) algorithms, they might achieve   better results, much sooner,  by applying hyper-parameter optimization to simple (but very fast)  algorithms. 

\subsection{Preliminaries}

Before beginning, just to say the obvious,
we note the experiments of this paper are based on two case studies. Hence, they do not show that {\em all} deep learners can be replaced by faster and simpler methods. 

That said, we would argue that this paper is at the very least arguing for a methodological change in how software analytics researchers report their deep learning results. Deep learners (or, indeed, any data mining results) should be compared to a simpler baseline method (in our case, feedforward networks) and also be adjusted via automatic tuning algorithms. The experience of this paper is that such a baseline + tuning analysis can lead to challenging and insightful results.

% \begin{tcolorbox}[colback=gray!10,colframe=gray!75!black,title=Glossary: Deep learning]
% \textbf{Perceptrons:} The earliest forms of neural networks \cite{rosenblatt1961principles}, consisting of multiple layers of connected nodes.

% \textbf{Feedforward neural network:} A slightly advanced model \cite{lecun2015deep}, which applies an ``activation function" at each node after performing a matrix multiplication of the weights with the inputs.

% \textbf{Activation function:} An arbitrary function applied at each layer of the network after matrix multiplication.

% \textbf{Architecture:} The arrangement of nodes and the connections between them.

% \textbf{Loss function:} The function (of the predictions and the target labels) tuned by the backpropagation \cite{rumelhart1985learning} ithmithm.
% \end{tcolorbox}

\section{Case Studies}
\label{sec:background}

Before going into algorithmic details,
this paper first presents the two domains that will be explored by those algorithms.

\subsection{Vulnerability Detection}
\label{sec:vuln}

Cyber attacks often rely on software vulnerabilities, i.e., unintentional security flaws in software that can be taken advantage of to obtain unauthorized access, steal data, etc. As of writing this paper, the Common Vulnerabilities and Exposures (CVE) database\footnote{\url{https://cve.mitre.org/}} contains over 165,000 records of vulnerabilities. This number only counts the \textit{registered} vulnerabilities, and not unknown (or ``zero-day") vulnerabilities. For the security of software systems and the data associated with them (for example, in SQL databases), it is critical that these vulnerabilities be discovered and patched. However, manually searching for vulnerabilities is a time-consuming task.

There are several existing solutions that attempt to automate this task \cite{viega2000its4, grieco2016toward, kim2017vuddy}. However, these rely on significant human effort. Specifically, they rely on the use of human-generated features, which can take time, and be expensive (since skilled human time is expensive). Moreover, these approaches tend to either have too many false negatives (i.e., missed vulnerabilities), or too many false positives (i.e., a ``learner'' that blindly marks non-vulnerable code as a vulnerability). These issues make these techniques less useful in practice.

\subsubsection{Algorithms for Vulnerability Detection}

To tackle these two problems, deep learning solutions have been recently proposed. \citet{li2018visualizing} propose VulDeePecker, a bidirectional LSTM \cite{hochreiter1997long} technique. From an external perspective, their approach takes in program segments, trains a deep learner, and then uses it to detect vulnerable code. Because this approach relies on training on the code to generate vector representations (which the network then uses to make predictions), it can be slow to run. \citet{zhou2019devign} propose Devign, which instead uses graph neural networks \cite{kipf2016semi} to detect vulnerabilities. A graph neural network takes in a graph input, and uses ``graph convolutions'' to extract hierarchical features. These features can then be used to make predictions in the later layers of the network. The authors of Devign experiment with several graph representations of source code, and recommend a composite version of their approach. Based on our literature review, we assert that this is the state-of-the-art approach for vulnerability detection.

However, deep learning approaches themselves can have issues. The major one is that deep learners can be slow to run \cite{jiang2018linear,le2011optimization,martens2010deep}. The primary reason for this is the use of more modern deep learning techniques such as the above mentioned bidirectional LSTMs. While these certainly have a lot of representational capacity, they suffer from having orders of magnitude more parameters than simpler, feedforward networks, and therefore take longer to optimize.

In this paper, we take a similar approach to VulDeePecker in that we use a deep learning technique to transform code into a vector representation, and then use our simple feedforward networks for prediction. However, unlike their approach, we use an off-the-shelf code-to-vector transformation tool, code2vec \cite{alon2019code2vec}. Because there is no training involved in using this model off-the-shelf, our runtimes are significantly faster, since only the feedforward networks need to be trained.

\subsection{Predicting Bugzilla Issue Close Time}\label{sec:tasks}

When programmers work on repositories, predicting  issue  close  time  has  multiple  benefits  for  the  developers,  managers, and stakeholders since it helps:
\begin{itemize}
\item
Developers prioritize work; 
\item
Managers allocate resources and improve consistency of release cycles; 
\item
Stakeholders understand changes in project timelines and budgets. 
\item
It is also useful to predict issue close time when an issue is created; e.g. to send a notification if it is  predicted that the current issue is an easy fix.
\end{itemize}
 We explore issue close time, for two reasons.
Firstly, it is a well
studied problem~\cite{lee2020continual,rees2017better,vieira2019reports,akbarinasaji2018predicting,guo2010characterizing,giger2010predicting,marks2011studying,Kikas16,habayeb2017use}.
Secondly, recent work has proposed a state-of-the-art deep learning approach to  issue close time prediction (see the   DeepTriage
deep learning systems from COMAD'19, described later in this paper~\cite{mani2019deeptriage}).

% The estimated time to close an issue is helpful for managers to assign priorities, for developers to design and refactor code accordingly, for end-users who are directly affected by the bug, and for stakeholders who have vested interests in the product itself:
% \begin{itemize}
% \item
% Although bugs have an assigned severity, this is not a sufficient predictor for the lifetime of the issue. For example, the author who issued the bug may be significant, if, for example, they are a significant contributor to the project.
% \item
% Alternatively, an issue deemed more \textit{visible} to end-users may be given higher priorities. It is therefore insufficient simply to consider the properties of the issue itself (i.e., the \textit{issue metrics}), but also of its environment (i.e., \textit{context metrics}). This is similar to recent work on how \textit{process metrics} are better defect predicting measures than \textit{product metrics} \cite{majumder2020revisiting}.
% \end{itemize}

\subsubsection{Traditional Algorithms for Predicting  Issue Close Time}

Most large software systems have a system to track bugs, or issues, in the product. These issues typically go through 
the same lifecycle, in which they transition across various states, including UNCONFIRMED and CLOSED, while also being assigned final states such as WONTFIX \cite{weiss2007long}.

To find prior work on predicting issue close time, we searched for papers in the last ten years (since 2010) in  Google Scholar using keywords ``bug fix time", ``issue close time", and ``issue lifetime". Then, we filtered them according to the criterion that they must be published in a top venue according to Google Scholar metrics Software Systems\footnote{\url{https://scholar.google.com/citations?view_op=top_venues&hl=en&vq=eng_softwaresystems}}.  Finally,
using engineering judgement, we added in systems that  were recommended by reviewers of a prior draft of this paper.
That search found several noteworthy systems:
\begin{itemize}
\item
\citet{guo2010characterizing} use logistic regression on a large closed-source project (Microsoft Windows), to predict whether or not a bug will be fixed. Using regression analysis, they   identified the factors that led to bugs being fixed or not fixed.
\item
\citet{giger2010predicting} use decision trees to predict the bug-fix time for Mozilla, Eclipse, and GNOME projects. They divided their target class into two labels: fast and slow, to get a binary classification problem, and used the area under the ROC curve (AUC) metric as their evaluation criteria.
\item
\citet{marks2011studying} also used decision trees, but instead, use an ensemble method, i.e., random forests, on Eclipse and Mozilla data. Their motivation for using random forests, apart from the better performance as compared to standard decision trees, is the ability to extract the relative importance of features in the input data. They report accuracy scores of 63.8\% and 67.2\% on the Mozilla and Eclipse repositories respectively.
\item
At MSR’16, Kikas, Dumas, and Pfahl~\cite{Kikas16}  built time-dependent models for issue close time
prediction using Random Forests with a combination of
static code features, and non-code features to predict issue
close time with high performance
\item
More recently, \citet{habayeb2017use} reported in   IEEE TSE'17 a prediction system based  on  hidden Markov chains. Like \citet{giger2010predicting}, they divided their target labels into fast and slow fix-times and experimented with different values of the number of hidden states of the hidden Markov model.
\end{itemize}
Based on the above, we assert that the two prior state-of-the-art non-neural
methods in area used random forests and logistic regression.
 Hence we will we use these two systems as part of the following study.
 
 \subsubsection{Deep Learning and  Issue Close Time}
As to deep learning and issue close time prediction, 
two contenders for ``state-of-the-art'' are DASENet~\cite{lee2020continual} and DeepTriage\cite{mani2019deeptriage}. The DASENet paper asserts that their algorithm
defeats DeepTriage but, after much effort, we could not reproduce that result\footnote{We found that the  reproduction package published with DASENet has missing files. We tried contacting the authors of that paper, without success.}. 
Hence, for this study, we use DeepTriage since:
\begin{itemize}
\item
It is a state-of-the-art deep learner   that performs for  lifetime prediction. 
\item
It has been very recently published (2019);
\item
Its reproduction package allowed us to run that code on our machines.
\item 
It uses datasets commonly used in the literature (Technical aside: we were tempted to use the dataset  provided by~\citet{vieira2019reports} for our deep learning baseline.
However, their lack of prior benchmarks meant we could not provide a comparison to demonstrate the efficacy of our approach.)
\end{itemize}
From a technical perspective, DeepTriage  is
\citet{mani2019deeptriage}'s extension of   bidirectional LSTMs with an ``attention mechanism''. A Long Short-Term Memory (LSTM) \cite{hochreiter1997long} is a form of recurrent neural network that has additional ``gate" mechanisms to allow the network to model connections between long-distance tokens in the input. Bidirectional variants of recurrent models, such as LSTMs, consider the token stream in both forward and backward directions; this allows for the network to model both the previous and the following context for each input token.  Attention mechanisms\cite{bahdanau2014neural} use learned weights to help the network ``pay attention" to tokens that are more important than others in a context.
Prior to running DeepTriage, its  authors recommend using  a standard set of preprocessing techniques: pattern matching to remove special characters and stack traces, tokenization, and and pruning the corpus to a fixed length. Beyond these steps, they rely on the deep learner to perform automated feature engineering.

% In our literature review (below), we  comment on the lack of comparison of deep learning to non-deep learning approaches. Accordingly, in our comparisons, we will also compare
% our work to the decision trees used by ~\citet{rees2017better}.
% In their study, they compare the effectiveness of cross-validation versus round robin approaches; tangentially, we use their results as a non deep learning benchmark; in addition, their dataset is multi-class and heavily skewed--this creates an additional challenge for deep learners, which are not tuned for metrics like recall and false alarm rate \cite{yedida2020improving}.

% In summary, we use two baseline results to compare against, which we believe are state-of-the-art and recent. These are listed below:

% \begin{itemize}
%     \item The results of Rees-Jones et al. \cite{rees2017better}, who use non-DL methods, and advocate for a round-robin approach over cross-validation.
%     \item The results of Lee et al. \cite{lee2020continual}, who demonstrate the results of the DeepTriage deep learner  on Eclipse, Firefox, and Chromium data mined from Bugzilla.
% \end{itemize}

% Therefore, we have a deep learning and a non-DL state-of-the-art results, across different datasets. For each comparison, we use the same datasets as the original authors, since both make their data publicly available.

\section{Algorithms}
Having discussed the domains we need to explore,
this paper now turns to how we will explore them.

For the purposes of exposition, we label divide this discussion
on these algorithms into three groups:
\begin{itemize}
\item Older style Feedforward Networks
\item Newer-style Deep Learners
\item Hyperparameter optimizers, which we use to tune
the parameters of Feedforward Networks and Deep learners
\end{itemize}
Note that the system we are calling SIMPLE is a combination
of Feedforward Networks and Hyperparameter Optimization.

\subsection{ Feedforward Networks} 
 
Feedforward neural networks~\cite{lecun2015deep}  apply a general ``activation function" 
%(rather than a threshold) 
at each node after performing the matrix multiplication of the weights with the inputs. These networks grew in popularity following the invention of the ReLU (rectified linear unit) function \cite{nair2010rectified}, $f(x) = \max(0, x)$, which significantly improved the results of neural networks. 
%This architecture is shown in Figure \ref{fig:dnn}. In contrast to perceptrons, feedforward networks typically (a) have fewer layers (b) use more sophisticated activation functions as opposed to a threshold (i.e., in Figure \ref{fig:dnn}, the way each $a^{[l]}_i$ is calculated is different). 
Specifically,  for a layer $i$, if the weight matrix is represented in matrix form as $W^{[i]}$, the \textit{bias} terms (the constants) are represented by $b^{[i]}$, and the values of the activation function are represented as $a^{[i]}$, then
$a^{[0]}  = X$ and
$z^{[i]}   = W^{[i]T}a^{[i-1]} + b^{[i]}$ and
       $a^{[i]}  = f(z^{[i]}) $ where $X$ is the input matrix.
     
There are several activation functions; for brevity, we only discuss the ones relevant in this study.
Following the advice of \citet{lecun2015deep}, 
for binary and multi-classification problems:
\begin{itemize}
\item
For the last layer of the network, this study uses 
{\em Sigmoid(x)}  $ = \frac{1}{1+e^{-x}}$
and 
{\em Softmax(x)} $  = \frac{\exp(x_k)}{\sum_{j=1}^{|x|} \exp(x_j)}$  respectively.
\item
For the other layers, we use  
{\em ReLU(x)}  $ = \max(0, x) $.
\end{itemize}

\subsection{  Deep Learning}
\label{sec:deeplearning}

\begin{figure}[!t]
\centering
\includegraphics[width=0.45\textwidth]{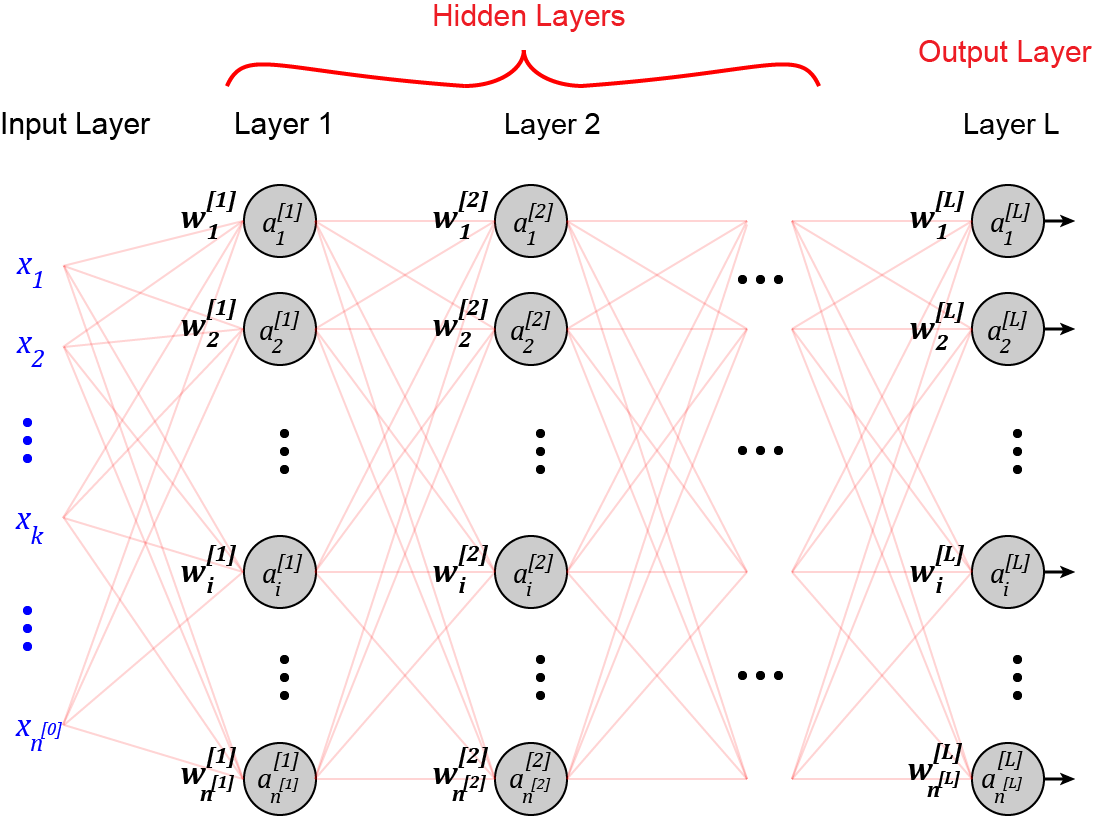}
\caption{Illustration of a neural net model. Feedforward networks, such as those used in SIMPLE, have far fewer hidden layers than deep learners.}    
\label{fig:dnn}
\end{figure}

For the rest of this paper, the following distinction will be important:
\begin{itemize}
\item
The algorithms DeepTriage  and VulDeePecker (used for issue close time and vulnerability defection, respectively)
are based on {\em new} neural network technology comprising extensive layers
of reasoning, where layer $i$ organizes the inputs offered to layer $i+1$.
\item
Our SIMPLE  method is based on {\em old} feedforward neural  networks which is a technology that dates back decades. At each node of these networks, the inputs are multiplied with weights that are learned, and then an activation function is applied. The weights are learned by the backpropagation algorithm \cite{rumelhart1985learning}.
 \end{itemize}
The difference between these approaches can be understood via Figure~\ref{fig:dnn}.
The  older
 methods use just a few layers while the ``deep" learners use many layers. Also, the older
 methods use a threshold function at each node, while feedforward networks typically use the ReLU function $f(x) = \max (0, x)$.

\subsection{ Hyperparameter Optimization} \label{sec:hyperparam-opt}

A common factor in all neural networks (feed-forward, deep learner, etc)
is the {\em architecture}  of the many layers of
neural networks ~\cite{goodfellow2016deep}. In deep learning terminology, an ``architecture" refers to the arrangement of nodes in the network and the connections between them, which dictates how the backpropagation algorithm updates the parameters of the model. Depending on the choice of the optimization algorithm (such as Adam ~\cite{kingma2014adam}) and the architecture used, the model also has several hyper-parameters, such as the number of layers, the number of nodes in each layer, and hyper-parameters of the optimization algorithm itself~\cite{brown2020language}. 

The selection of appropriate hyper-parameters is something of a black art. Hence there exists a whole line of research called hyper-parameter optimization that explores automatic methods for finding these values.

For this study, we consider using two such optimizers: {\em TPE} (tree-structured Parzen estimators) from
 Bergstra et al.~\cite{bergstra2012random,bergstra2011algorithms}
and {\em DODGE} from Agrawal et al.~\cite{agrawal2019dodge,agrawal2020simpler}:
\begin{itemize}
\item
 {\em TPE} is a candidate hyper-parameter tuner since
a December 2020 Google Scholar search for ``Hyper-parameter optimization'' 
 reported that    papers by 
 Bergstra et al.~\cite{bergstra2012random,bergstra2011algorithms} on   TPE   optimization have more citations (2159 citations  and 4982 citations\footnote{The nearest other work was a 2013 paper by Thornton et al. on Auto-WEKA~\cite{thornton2013auto} with 931 citations.}) that any other paper in this arena.
 \item
  {\em DODGE} is another  candidate  hyper-parameter since, unlike {\em TPE}, it has been extensively tested on SE data sets.
In 2019, Agrawal et al.~\cite{agrawal2019dodge}  reported that for a range of SE problems (bad small detection, defect prediction, issue severity prediction) learners tuned by DODGE out-perform    prior state-of-the art results
(but a missing part of their analysis is that they did not study
deep learning algorithms, hence, this paper).
\end{itemize}
How to choose between these algorithms?
In 2021, Agrawal et al.~\cite{agrawal2020simpler} showed that {\em DODGE} is preferred over {\em TPE}
for   ``intrinsically simple'' data sets.
 \citet{levina2004maximum}
  argue that many datasets     embedded in high-dimensional spaces can be compressed without significant information loss.
  They go on to say that a simple linear transformation like Principal Components Analysis (PCA) \cite{pearson1901liii} is insufficient, as the lower-dimensional embedding of the high-dimensional points are not merely projections.
Instead, \citet{levina2004maximum} propose a method that computes the intrinsic dimensionality by counting the number of points within a distance $r$ while varying $r$. For notes on that computation, see 
Table~\ref{dim}

\begin{table}[!t]
\small
\caption{Feedforward networks are controlled by these hyper-parameters.}\label{tunings}
\begin{tabular}{|p{.95\linewidth}|}\hline \\
\textbf{Preprocessors:}
\begin{itemize}
\item StandardScaler : i.e. 
all input data set numerics are adjusted
to $(x-\mu)/\sigma$.
\item MinMaxScaler (range = (0, 1)):  i.e.  scale each feature to $(0,1)$.
\item Normalizer (norm =  randchoice([`l1', `l2',`max'])): i.e. normalize  to a unit norm.
 \item MaxAbsScaler (range = (0, 1)): scale each feature by its maximum absolute value
\item Binarizer (threshold =   randuniform(0,100)), i.e., divide  variables on some  threshold
\end{itemize}\\\hline \\
\textbf{Hyper-parameters:}
\begin{itemize}
    \item Number of layers
    \item Number of units in each layer
    \item Batch size (i.e., the number of samples processed at a time)
\end{itemize} \\\hline
\end{tabular}
\end{table}

Intrinsic dimensionality (which we will denote as $D$) 
can be used to select an appropriate hyper-optimization strategy.
Agrawal et al.~\cite{agrawal2020simpler}.
experiments show that  {\em DODGE}  beasts {\em TPE} for low dimensional data 
(when $D<8$) while
{\em TPE} is the preferred algorithm for more complex data. 
 
\begin{table}[!t]
\small
\caption{Notes on intrinsic dimensionality.}\label{dim}
\begin{tabular}{|p{.95\linewidth}|}\hline
Before presenting the mathematics of the \citet{levina2004maximum} measure, we offer a little story to explain the  intuition behind this  measure

Consider a brother and sister who live in different parts of town. The sister lives alone, out-of-town, on a road running north-south with houses only on one side of the street.   Note that if this sister tries to find company by walking:
 \begin{itemize}
 \item
 Vertically up or down;
 \item Or east or west
 \end{itemize} then she will meet no one else. But if she walks north or south, then she might  find company. That is,  the humans in that part of town  live in a one-dimensional space (north-south). 
Meanwhile, the brother lives  downtown in the middle of a large a block of flats that is also oriented north-south.  The brother is ill-advised to  walk east-west since then they will fall off a balcony. On the other hand, if  he :
\begin{itemize}
 \item
 Climbs  up or down one storey 
 \item Or walks to the neighboring flats north or south
 \end{itemize}
then the brother might meet other people.  That is to say, the humans in that block of flats  effectively live in a two-dimensional space (north-south and up-down).

To compute Levina's intrinsic dimensionality, we create a 2-d plot where the x-axis shows $r$; i.e. how far we have walked away from any instance
and the y-axis show $C(r)$ which counts how  many more people we have meet after walking some distance $r$ way from any one of $n$ instances:
\[
y = C(r) = \frac{2}{n(n-1)}\sum\limits_{i=1}^n \sum\limits_{j=i+1}^n I\left[\lVert x_i, x_j \rVert < r \right]
\]
The maximum slope of  $\ln C(r)$ vs. $\ln r$ is then reported as the intrinsic dimensionality.
Note that
$I[\cdot]$ is the indicator function (i.e., $I[x] = 1$ if $x$ is true, otherwise it is 0);
$x_i$ is the $i$th sample in the dataset.
Note also that, 
as shown by \citet{aggarwal2001surprising}, at higher dimensions the distance calculations should use the $L_1$ norm, i.e., $\sum \lvert x_i \rvert$ rather than the $L_2$ norm, i.e., $\sqrt{\sum x_i^2}$.\\\hline
\end{tabular}

\end{table}
Using the calculation methods of Agrawal et al.~\cite{agrawal2020simpler},
we find that  for our data:
\[
\mathit{D(Firefox, \;Chromium, \;Eclipse)} = \{2.1, \;1.95,\; 1.9\}
\]
From this, we make two observations.
Firstly,
in a result that may not have surprised  Levina et al.,
this data from Firefox, Chromium, Eclipse  can be compressed w to just a few dimensions.
Secondly,
all our data can be found below the  $D<8$ threshold proposed by Agrawal et al.~\cite{agrawal2020simpler}.
Hence, for this study, we use  {\em DODGE}.
 
 Compared to other hyper-parameter tuners, 
 {\em DODGE} is a very simple algorithm that runs in two steps:
 \begin{enumerate}
\item
During an  initial random step, {\em DODGE} selects
hyper-parameters at random from Table~\ref{tunings}.
Each such tuning is used to configure a learner. The value
of that configuration is then assessed by applying that learner
to a data set.
If ever a NEW result has performance scores near an OLD result, then a ``tabu''  zone is created around OLD and NEW configurations that subsequent random searches avoid that region of    configurations.
\item
In the next step,  {\em DODGE}   selects configurations via a 
 binary chop of the tuning space. Each chop moves in the bounds for numeric choices
by half the distance from most distant value to the value that produced the ``best'' performance. For notes on what ``best'' means, see \S\ref{perform}.
\end{enumerate} 
Agrawal et al. recommend less than 50 evaluations for each of {\em DODGE}'s two stages. Note that this is far less than other hyper-parameter optimizations strategies.
To see that, consider another
hyper-parameter optimization approach based on  genetic algorithms that mutate $P$ individuals over $G$ generations (and between each generation, individuals give ``birth'' to new individuals
 by crossing-over   attributes from two parents). Holland~\cite{john1992holland} recommends P=G=100 as useful defaults for genetic algorithms. Those default settings implies that a standard genetic algorithm tuner would
 require $100*100=10,000$ evaluations.

Note that we also considered tuning DeepTriage,  but that proved impractical:
\begin{itemize}
\item
The DeepTriage learner used in this study  can take up to  six CPU hours to learn
one model from the  issue close time data.
When repeated for 20 times (for statistically validity) over our (15) data sets, that  means
that using DODGE (using 42 evaluations)
on DeepTriage would require over 8 years of CPU time.
\item
On the other hand,   with 20 repeats over our datasets, DODGE with feedforward networks terminated in 26 hours;  i.e. nearly 2,700 times faster than tuning  DeepTriage.
\end{itemize}

\section{Experimental Methods}
\label{sec:method}

\subsection{Methods for Issue close time prediction}

This section discusses how we comparatively evaluate different ways to do issue close time prediction. We explore three learners:
\begin{itemize}
\item[L1:]
DeepTriage:
a state-of-the-art    deep learner from COMAD'19~\cite{mani2019deeptriage};
\item[L2:]
Our  SIMPLE neural network learner, described in \S\ref{simpler};
\item[L3:]
Non-neural approaches: random forest from ~\citet{marks2011studying}, and logistic regression from \citet{guo2010characterizing} (we present the better of the two results, where ``better'' is defined via the statistical methods of \S\ref{sec:stats}).
%\item[L4:]
%A non-neural approach  from IEEE TSE'17 based on Markov chains~\cite{habayeb2017use}
\end{itemize}
These learners will be studied twice:
\begin{itemize}
\item[S0:] Once, with the default off-the-shelf settings for  learners control parameters;
\item[S1:] Once again, using the settings found after some automatic tuning.
\end{itemize}
The original research plan was to present six sets of results:
\begin{center}
planned = \{L1,L2,L3\} * \{S0,S1\}
\end{center}
However, as noted below, the tuning times from DeepTriage were so slow that we could only report five results:
\begin{center}
actual =  (\{L1\} * \{S0\}) + (\{L2,L3\} * \{S0,S1\})
\end{center}

\subsection{Methods for Vulnerability Detection}

For vulnerability detection, we use source code as the starting point for our approach. The first step is to convert the source code into a vector representation. For this, we use the code2vec method of \citet{alon2019code2vec}. Specifically, inspired by the Attention mechanism \cite{bahdanau2014neural,vaswani2017attention}, they propose a ``Path-Attention'' framework based on paths in the abstract syntax tree (AST) of the code. However, the two systems that we study (ffmpeg and qemu) are written in C++, while code2vec was initially built for Java code. To our benefit, code2vec uses an intermediate AST representation as its input, which we convert using the astminer toolkit\footnote{\url{https://github.com/JetBrains-Research/astminer}}. Having done that, we then use code2vec to create vector representations of our two software systems. Next, we reduce the dimensionality of these vectors using an autoencoder, an encoder-decoder architecture \cite{badrinarayanan2017segnet} that performs non-linear dimensionality reduction. Finally, we perform random oversampling to handle class imbalance.

We emphasize here that this step is \textit{preprocessing}, not training. Any software analytics solution that can be applied effectively in the real world must use source code as input; however, machine learning models expect vector inputs. Therefore, a preprocessing step is necessary to bridge this representation gap between the raw source code and the input to the machine learning system. For example, \citet{li2018vuldeepecker} use a bidirectional LSTM model to extract vectors, and append a \textit{Dense} layer to this deep learner to make predictions. Training end-to-end in this manner has the advantage of simplicity, but comes at a computational cost since each training step also trains the preprocessor. By decoupling these two parts, we allow for training the preprocessor once (per software system) and then using the actual learner (in our case, the feedforward network) to make predictions.

We train our feedforward networks in a straightforward manner. We train for 200 epochs using the Adam optimizer with default settings. We perform hyper-parameter optimization using DODGE, for 30 iterations as recommended by its authors.

\subsection{Data for Issue close time prediction}\label{data}

To obtain a fair comparison with the prior state-of-the-art, we use the same data as used
in the prior study (DASENet) \cite{lee2020continual}.  
One reason to select this baseline is that  we were able to obtain the data used in the original study (see our reproduction package)
and,   therefore, were able to obtain results comparable to prior work.   For a summary of that data,
see Table~\ref{tab:bugzilla-data}.

% \citet{rees2017better} gather data from GitHub and JIRA, which were in JSON form with commit data, issues, and contributor details. The authors follow the feature engineering steps of \citet{kikas2016using}, but do not collect temporal data, which led to only seven features. Issues which were not yet closed at the time of data collection were removed. Then, they start feature engineering with the correlation-based feature selection (CFS) algorithm \cite{hall1999correlation}, which works on feature subsets rather than individual features. CFS is based on the principle that a ``good" subset of features is one that is highly correlated with the target label, but has the property that the individual features comprising the subset are weakly correlated with each other. CFS performs a best-first search to find the best subset of features. Specifically, for a subset of $k$ features, whose correlation is denoted by $\rho$ and the correlation between the features is $\sigma$, the value optimized is given by $\frac{k\rho}{\sqrt{k+(k-1)\sigma}}$.

For the comparison with the \citet{mani2019deeptriage} study, the data was collected from Bugzilla for the three projects: Firefox, Chromium, and Eclipse:
\begin{itemize}
\item
To collect that data, \citet{mani2019deeptriage} applied
  standard text mining preprocessing   (pattern matching to remove special characters and stack traces, tokenization, and and pruning the corpus to a fixed length).
  \item
  Next, the activities of each day were collected into ``bins", which contain metadata (such as whether the person was the reporter, days from opening, etc.), system records (such as labels added or removed, new people added to CC, etc.), and user activity such as comments.
  \item
  The metadata can directly be represented in numerical form, while the user and system records are transformed from text to numerical form using the word2vec \cite{mikolov2013efficient, mikolov2013distributed} system. These features, along with the metadata, form the input to the DeepTriage \cite{mani2019deeptriage} system and our feedforward learners for comparison. 
\end{itemize}
In the same manner as prior work using the Bugzilla datasets, we discretize the target class into 2, 3, 5, 7, and 9 bins (so that each bin has roughly the same number of samples). This yields datasets that are near-perfectly balanced (for example, in the Firefox 2-class dataset, we observed a 48\%-52\% class ratio).

\begin{table}[!t]
    \centering
    \caption{Issue close time prediction data. From~\citet{lee2020continual} study. Note that because of the manner of data collection, i.e., using bin-sequences for each day for each report, there are many more data samples generated from the number of reports mined.}
    \small
    \begin{tabular}{llrrr}
    \toprule \\
    \textbf{Project} & \textbf{Observation Period} & \textbf{\# Reports} & \textbf{\# Train} & \textbf{\# Test} \\ \midrule
        Eclipse & Jan 2010--Mar 2016 & 16,575 & 44,545 & 25,459 \\
        Chromium & Mar 2014--Aug 2015 & 15,170 & 44,801 & 25,200 \\
        Firefox & Apr 2014--May 2016 & 13,619 & 44,800 & 25,201 \\
        \bottomrule
    \end{tabular}
    \label{tab:bugzilla-data}
\end{table}

\begin{table}
    \centering
    \caption{Vulnerability Detection datasets, from the Devign \cite{zhou2019devign} paper. VFCs = Vulnerability-Fixing Commits.}
    \label{tab:vuln-data}
    \begin{tabular}{llll}
        \toprule
        \textbf{Project} & \textbf{Total commits} & \textbf{VFCs} & \textbf{Non-VFCs} \\
        \midrule
        qemu & 11,910 & 4,932 & 6,978 \\
        ffmpeg & 13,962 & 5,962 & 8,000 \\
        \bottomrule
    \end{tabular}
\end{table}

\subsection{Data for Vulnerability Detection}

For vulnerability detection, we use the datasets provided by \citet{zhou2019devign}. However, although the authors test their approach on four projects, only two are released: ffmpeg and qemu. These are two large, widely used C/C++ applications: ffmpeg is a library that handles audio and video tasks such as encoding; qemu is a hypervisor. To collect this data, the authors collected vulnerability-fixing commits (VFCs) and non-vulnerability-fixing commits (non-VFCs) using (a) keyword-based filtering of commits based on the commit messages (b) manual labeling. Then, vulnerable and non-vulnerable functions are extracted from these commits. The authors use Joern \cite{yamaguchi2014modeling} to extract abstract syntax trees, control flow graphs, and data flow graphs from these functions.

In total, for qemu, the authors collected 11,910 commits, of which 4,932 were VFCs and 6,978 were non-VFCs. For ffmpeg, the authors collected 13,962 commits, of which 5,962 were VFCs and 8,000 were non-VFCs. This data is summarized in Table \ref{tab:vuln-data}.

\subsection{Tuning the  SIMPLE Algorithm}\label{simpler}
Our SIMPLE algorithm is shown in Algorithm \ref{alg:simple}.

Table~\ref{tunings} shows the parameters that control the
feedforward network used by SIMPLE.

One issue with any software analytics paper is how researchers
decide on the ``magic numbers'' that control their learners (e.g. Table~\ref{tunings}).
 
In order to make this paper about simpler neural feedforward networks versus deep learning
(and not about complex methods for hyper-parameter optimization), we selected the controlling hyper-parameters for the feedforward networks using     hyper-parameter optimization.

\begin{algorithm}[!t]
\footnotesize
    Set random number seed\; 
    \For{20 times}{
   Shuffle data\;
   Set train, test =  70\%,30\% splits of the data\;
%   \tcc{Feature engineering with BUFFER} 
%   \For{eg in train}{
%       \If{eg.class== target}{   
%          $\epsilon = 0.1$\;
%           \If{ distance(from eg to closest instance of another class) $\le$ $2*\epsilon$}{
%                  \tcc{apply the BUFFER algorithm} 
%                  at distance $4\epsilon$, $2\epsilon$, $\epsilon$  CREATE $1,2,4$ extra examples}}}
%   \tcc{Feature engineering with SMOTE} 
%     {\em imbalance} = 0.2\;
%   \While{ abs(percent(class1) - percent(class2)) $>$ imbalance}{
%       \tcc{apply the SMOTE algorithm}  
%       Remove any majority example (selected at random)\;
%       CREATE one minority example} 
     \tcc{Learning}  
     Apply a feedforward neural network.;
     On the training data,   tune the hyper-parameters of Table~\ref{tunings} using DODGE (see \S\ref{sec:hyperparam-opt}).\;
    Take the best model found from the training data, apply it to the test   data\;
    Report performance scores on the test data. \;}
    \caption{SIMPLE}
    \label{alg:simple}
\end{algorithm}

\subsection{Performance Metrics}\label{perform}
 
%\citet{lee2020continual} and \citet{rees2017better} use many metrics for evaluating their models.

% For each dataset, we use the same metrics as our baselines for a fair comparison against the prior state-of-the-art results. In the subsequent text, we use $TP$ to mean true positives, $TN$ to mean true negatives, $FP$ to mean false positives, and $FN$ to mean false negatives.

% \subsubsection{Rees-Jones et al.}

% \citet{rees2017better} use the following four metrics.

% \begin{itemize}
%     \item \textbf{Recall} measures the fraction of positive samples identified by the classifier.
%   \mbox{ $\textbf{pd} = \mathit{TP}/(TP + FN)$}

%     \item \textbf{False alarm rate} is the fraction of negative samples falsely predicted as positive by the classifier.
    
%     \mbox{
%         $\textbf{pf} = \mathit{FP/(FP + TN)}$
%     }
    
%     \item \textbf{Precision} is the fraction of modules correctly predicted as positive.
    
%     \mbox{
%         $\textbf{prec} = \mathit{TP/(TP + FP)}$
%     }
    
%     \item \textbf{F-1 score} is the harmonic mean of recall and precision.
% \end{itemize}

% To these four metrics, we also add the distance to heaven (d2h) metric, which is defined as:
% $ \textbf{d2h} = \sqrt{\frac{(1 - \text{pd})^2 + \text{pf}^2}{2}}$. We use d2h because it describes the balance between modules correctly identified as positive (recall) without predicting everything as a positive class (false alarm rate).

% For d2h and false alarm rate, \textit{smaller} values are \textit{better}. For the rest, larger values are better.

%\input{mitch}
% \subsubsection{Lee et al.}

 Since we wish to compare our approach to prior work, we take the methodological step of adopting the same performance scores as that seen in prior work.\citet{lee2020continual} use the following two metrics in their study:

\begin{itemize}
    \item \textbf{Accuracy} is the percentage of correctly classified samples. If TP, TN, FP, FN are the true positives, true negatives, false positives, and false negatives (respectively), then {\em accuracy} is $\mathit{(TP+TN)/(TP+TN+FP+FN)}$.
    
    \item \textbf{Top-2 Accuracy}, for multi-class classification, is defined as the percentage of samples whose class label is among the two classes predicted by the classifier as most likely. Specifically, we predict the probabilities of a sample being in each class, and sort them in descending order. If the true label of the sample is among the top 2 classes ranked by the classifier, it is marked as ``correct".
\end{itemize}

Additionally, for vulnerability detection, \citet{zhou2019devign} use \textbf{F1-score} as their metric, which is defined as follows. Let \textit{recall} be defined as the fraction of true positive samples that the classifier correctly identified, and \textit{precision} be the fraction of samples classified as positive, that were actually positive. That is,

\begin{align*}
    \mathrm{Recall} &= \frac{TP}{TP + FN} \\
    \mathrm{Precision} &= \frac{TP}{TP + FP}
\end{align*}

Then F1-score is the harmonic mean of recall and precision, i.e.,

\[
    \mathrm{F1} = \frac{2 \cdot \mathrm{precision} \cdot \mathrm{recall}}{\mathrm{precision} + \mathrm{recall}}
\]

% In other software analytics work, other evaluation measures
% are used such as false alarm. Previously,
% we have critiqued those measures saying that they can have issues
% with data sets where one class is far more frequent than another~\cite{menzies2007problems}. In the issue lifetime prediction dataset, those concerns do not apply since, as discussed in \S\ref{data}, the pre-processing of our data ensures that all our classes occur at equal ratios. Hence, the measures shown above (Accuracy and Top-2 Accuracy) are sufficient. 

\begin{table*}[!t]
    \centering
    \caption{Results on BugZilla data used in prior deep learning state of the art. The target label is discretized into a different number of classes (columns) as in the prior work. \colorbox{gray!20}{\textbf{Dark cells}} indicate statistically better performance.\\ \textbf{Key:} \textbf{DT} = DeepTriage \cite{mani2019deeptriage}; 
    \textbf{NDL-T} = best result of untuned non-neural methods; i.e. best of logistic regression~\cite{guo2010characterizing} and random forests~\cite{marks2011studying}; 
    \textbf{NDL+T} = best  of DODGE-tuned non-neural methods; i.e.  \textbf{NDL-T} plus tuning; 
    \textbf{FF} = untuned feedforward network; i.e Algorithm~1, without tuning; 
    \textbf{SIMPLE} = SIMPLE i.e.  \textbf{FF} plus tuning;  $T_k$ = Top-k accuracy;}
    \label{tab:bugzilla}
    \small
    \begin{tabular}{l|l|l|ll|ll|ll|ll}
    \toprule
        \textbf{Project} & \textbf{Model} & \textbf{2-class} & \multicolumn{2}{c}{\textbf{3-class}} & \multicolumn{2}{c}{\textbf{5-class}} & \multicolumn{2}{c}{\textbf{7-class}} & \multicolumn{2}{c}{\textbf{9-class}} \\ \midrule
         &  & $T_1$ & $T_1$ & $T_2$ & $T_1$ & $T_2$ & $T_1$ & $T_2$ & $T_1$ & $T_2$ \\ \midrule
        \multirow{5}{*}{Firefox} & DT & 67 & 44 & 78 & 31 & 58 & 21 & 39 & 19 & 35 \\ 
        & NDL-T & \gray{70} & 43 & 64 & 30 & 42 & 18 & 30 & 18 & 30 \\
        & NDL+T & 68 & 47 & 79 & 34 & 61 & 25 & 45 & 21 & 39 \\
        & FF & \gray{71} & 49 & 82 & 37 & 63 & 26 & 47 & 23 & 41 \\
         & SIMPLE & \gray{70} & \gray{53} & \gray{86} & \gray{39} & \gray{67} & \gray{37} & \gray{61} & \gray{25} & \gray{45} \\ \midrule
        \multirow{5}{*}{Chromium} & DT & 63 & 43 & 75 & 27 & 52 & 22 & 38 & 18 & 33 \\ 
        & NDL-T & 64 & 35 & 56 & 23 & 36 & 15 & 27 & 15 & 28 \\
        & NDL+T & 64 & 49 & 79 & 30 & 56 & 26 & 42 & 23 & 40 \\
        & FF & 65 & 53 & 82 & 35 & 60 & 27 & 45 & 26 & 42 \\
         & SIMPLE & \gray{68} & \gray{55} & \gray{83} & \gray{36} & \gray{61} & \gray{29} & \gray{48} & \gray{28} & \gray{45} \\ \midrule
        \multirow{5}{*}{Eclipse} & DT & 61 & 44 & 73 & 27 & 51 & 20 & 37 & 19 & 34 \\ 
        & NDL-T & 66 & 33 & 54 & 23 & 38 & 16 & 29 & 16 & 29 \\
        & NDL+T & 65 & 52 & 81 & 30 & 56 & 27 & 44 & 27 & 42 \\
        & FF & 66 & 54 & 81 & 32 & 59 & \gray{30} & \gray{47} & 30 & 46 \\
         & SIMPLE & \gray{69} & \gray{56} & \gray{84} & \gray{35} & \gray{62} & \gray{31} & \gray{48} & \gray{33} & \gray{49} \\ \bottomrule
    \end{tabular}
\end{table*}

\begin{table}
    \centering
    \caption{Vulnerability detection results.}
    \label{tab:vuln}
    { \small
    \begin{tabular}{lll}
        \toprule
        \textbf{Project} & \textbf{Model} & \textbf{F1-score}  \\
        \midrule
        \multirow{5}{*}{qemu} & NDL-T & 59 \\
        & NDL+T & 45 \\
        & FF & 51 \\
        & SIMPLE & \gray{73}  \\
        & Devign & \gray{73} \\
        \midrule
        \multirow{5}{*}{ffmpeg} & NDL-T & 52 \\
        & NDL+T & 52 \\
        & FF & 57 \\
        & SIMPLE & 67  \\
        & Devign & \gray{74}  \\
        \bottomrule
    \end{tabular} }
\end{table}

\subsection{Statistics}
\label{sec:stats}

Since some of our deep learners are so slow to execute,
 one   challenge in these results is to compare the results of a very slow system versus a very fast one (SIMPLE) where the latter can be run multiple times while it is impractical
to repeatedly run the former.
Hence,
for our definition of ``best'',  we will  compare
one result of size $|N_1|=1$ from the slower learner (DeepTriage) to a sample of $|N_2| =20$
results from the other.

Statistically, our evaluation of these results requires a check if   one results is less than a ``small effect'' different to the central tendency of the other population. 
For that statistical task,
~\citet{rosenthal1994parametric} says   there are two ``families'' of methods:  the $r$ group that is based on the Pearson correlation coefficient; or the $d$ family that is based on absolute differences normalized by (e.g.) the size of the standard deviation. \citet{rosenthal1994parametric} comment that ``none is intrinsically better than the other''. 
Hence, the most direct method is utilized in our paper. Using a $d$ family method, it can be concluded that one distribution is the same as another if their mean value differs by less than Cohen's delta ($d$*standard deviation).

{\footnotesize \begin{equation}\label{eq:cohen}
d=\mathit{small\; effect} = 0.3*\sqrt{\frac{\sum_i^x(x_i- ({\sum}x_i/n))^2}{n-1}}\end{equation}}
i.e., 30\% of the standard deviation of the $N_2$ population.

\section{Results}\label{results}

In this section, we discuss our results by answering two research questions:

\textbf{RQ1.} {\em  Does   ``Old but Gold''  hold for issue lifetime prediction?}

\textbf{RQ2.} {\em Does  ``Old but Gold''    hold for vulnerability detection?}

\subsection{RQ1: Issue lifetime prediction}

In this section, we discuss the answer to \textbf{RQ1}, which was, ``Does the \textit{Old but Gold} hypothesis hold for issue lifetime prediction?''

In Table~\ref{tab:bugzilla},  best results
are indicated by the 
\colorbox{gray!20}{\textbf{gray cells}}.  The columns of that
table describe how detailed are our time predictions. A column labeled $k$-class means that the data was discretized into $k$ distinct labels, as done in prior work (see \citet{lee2020continual} for details).

Recall that cells are in gray if the are statistically significantly better. In all cases, SIMPLE's results were (at least) as good as anything else.
Further, once we start exploring more detailed time divisions (in the 3-class, 5-class, etc problems) then SIMPLE is the stand-out best algorithm.
 
Another thing we can say about these results is that SIMPLE is much faster than other approaches. The above results took $\approx$ 90 hours to generate, of which 9 hours was required for SIMPLE (for 20 runs, over all 15 datasets) and 80 hours were required for the deep learner (for 1 run, over all 15 datasets). Recall that if we had also attempted to tune the deep learner, then that runtime would have exploded to six years of CPU.

From this discussion, we conclude RQ1 as follows:

\begin{blockquote}
    The ``Old but Gold'' hypothesis holds for issue lifetime prediction.
\end{blockquote}

\subsection{RQ2: Vulnerability detection}

In this section, we discuss the answer to \textbf{RQ2}, which was, ``Does the effect hold for vulnerability detection''?

Table \ref{tab:vuln} shows our results for vulnerability detection. While our data is limited (in that we could only use the two datasets released by the authors of \cite{zhou2019devign}), the data we do have suggests that SIMPLE can perform as well as Devign. In the case where SIMPLE lost, the difference was small (7\%). Therefore, we recommend the more complex deep learner when that 7\% is justified by domain constraints (e.g., a highly safety-critical system); however, a pragmatic engineering case could be made that the difference is marginal and negligible. We postulate that the slightly better performance of Devign is due to the superior preprocessing done by the multiple deep learning layers used by their approach, which allows for rich feature extraction and superior performance. That said, we argue that our approach runs faster than their sophisticated technique. While we could not reproduce their results (since their code is not open source), our approach takes 205 seconds on average, while their approach runs overnight\footnote{For their runtime, we contacted the authors, who reported that ``it ran overnight on their machines''.}.

Our conclusion is that:

\begin{blockquote}
    For vulnerability detection, the ``Old but Gold'' hypothesis worked for half the data sets studied here. 
\end{blockquote}

These results mean that we cannot unequivocally advocate simple methods for vulnerability detection. But then neither can these advocate for the use of deep learning for vulnerability prediction.  In our view,  these results strongly motivate the need for further study in this area
(since, if simpler methods do indeed prevail fro vulnerability detection, then this would
simplify research into   pressing current issues of software security).

\section{Threats to Validity}
\label{sec:threats}

\textbf{Sampling bias:} As with any other data mining paper, it is important to discuss sampling bias. We claim that this is mitigated by testing on 3 large SE projects over multiple discretizations, and demonstrating our results across all of them. Further, these datasets have been used in prior work that have achieved state-of-the-art performance recently. Nevertheless, in future work, it would be useful to explore more data. 

\textbf{Learner bias:} Our learner bias here corresponds to the choice of architectures we used in our deep learners. As discussed above, we chose the architectures based on our reading of ``standard DL'' from the literature. While newer architectures may lead to better results, the crux of this paper was on how simple networks suffice. Therefore, we maintain that the intentional usage of the simple, feedforward architecture was necessary to prove our hypothesis.

\textbf{Evaluation bias:} We compared our methods using top-1 and top-2 accuracy scores, consistent with prior work. These metrics are valid since the method the classes were discretized (as discussed in prior work) lends to equal-frequency classes. We further reduce the evaluation bias by running our experiments 20 times for each setup, and using distribution statistics, i.e., the Scott-Knott test, to check if one setup is significantly better than another.

\textbf{Order bias:} This refers to bias in the order in which data elements appear in the training and testing sets. We minimize this by running the experiment 20 times, each with a different random train-test split.

\textbf{External validity:} We tune the hyper-parameters of the neural network using DODGE, removing external biases from the approach. Our baseline results are based on the results of Montufar et al. ~\cite{montufar2014number}, which has been evaluated by the deep learning community. We also compare our work to non-deep learning methods, both with and without tuning by DODGE, to provide a complete picture of the performance of our suggested approach in relation to prior work and other learners.

\begin{figure}[!b]
    \centering
    \includegraphics[width=1\linewidth]{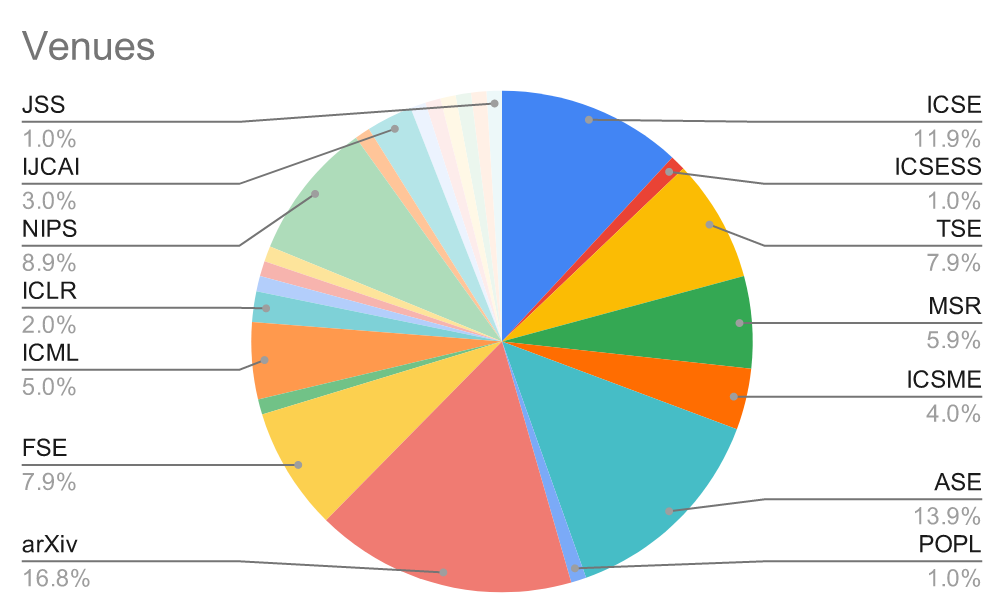}
    \caption{The distribution of papers across venues}
    \label{fig:venues}
\end{figure}

\begin{figure*}[!t]
    \centering
    \includegraphics[width=\textwidth]{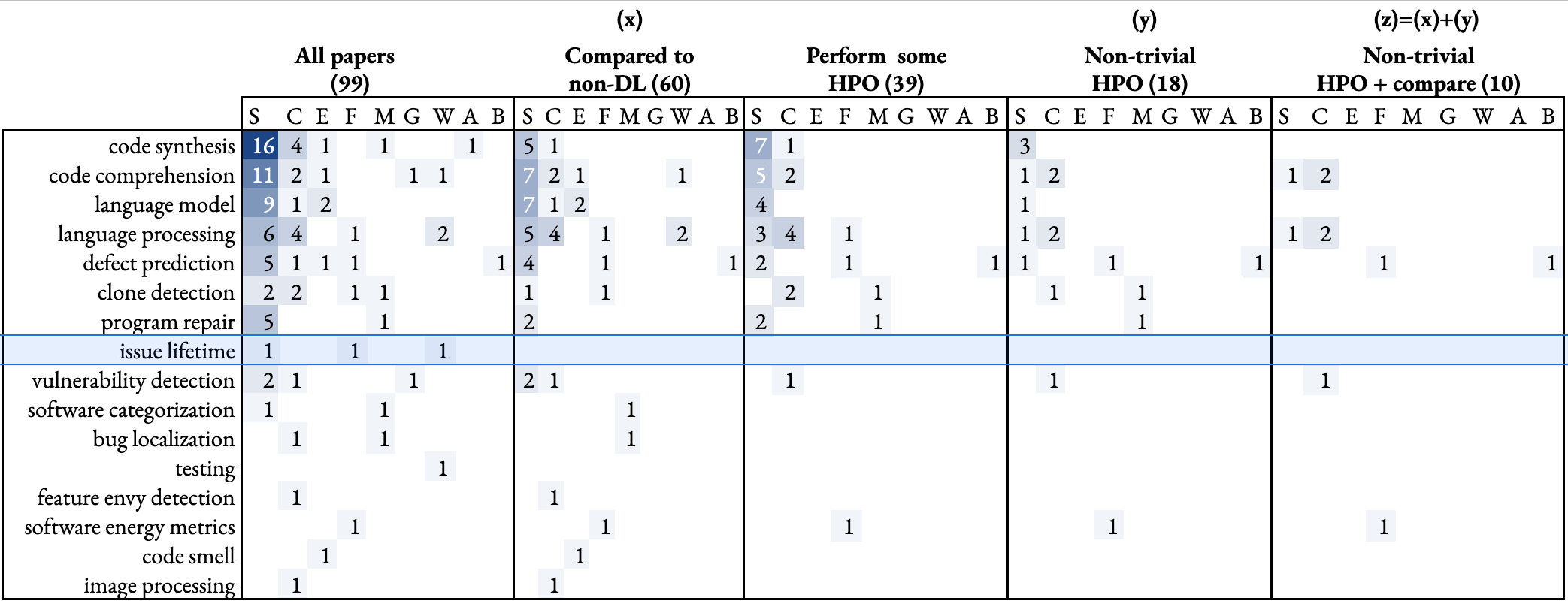}
    \caption{A summary of our literature review of deep learning methods in SE. The blue row denotes the DeepTriage system
    used in this paper.
    %plus a comparison against non-DL methods. \textbf{(a)} All papers in our literature review. \textbf{(b)}  Papers that compare their methods versus some non-DL method. \textbf{(c)} Papers performing hyper-parameter tuning.\textbf{(d)}  Papers performing non-trivial hyper-parameter tuning (i.e., not  using deprecated grid search~\cite{bergstra2012random} and using a hold-out set to assess the tuning before going to a separate test set). \textbf{(e)} Papers that satisfy both criteria (b) and (d).
  \textbf{Legend:}  \textbf{A} =    attention mechanism,
\mbox{\textbf{B} =    deep belief network},
\mbox{\textbf{C} =    convolutional networks},
\mbox{\textbf{E} =    embedding},
\mbox{\textbf{F} =    feedforward networks} (which includes traditional perceptrons~\cite{rosenblatt1961principles} \cite{mcculloch1943logical})
\mbox{\textbf{G} =    graph networks},
\mbox{\textbf{M} =    misc (i.e. some particular architecture invented by the author, used in one paper)},
\mbox{\textbf{S} =    sequence},
\mbox{\textbf{W} =    word2vec}. 
For a list of the papers shown in the right-hand-side column, see Table~\ref{tab:papers}.
}
    \label{fig:heatmap-total}
\end{figure*}

\section{Literature Review: deep learning in SE}

Using a  literature review,
this section argues that
the  issue raised in this paper
(that   researchers
seen rush to use the latest methods from deep learning literature, without baselining them against simpler)
is widespread in the software analytics literature.

To understand how deep learning are used in SE, we  performed  the following steps.

\begin{itemize}
    \item \textbf{Seed:} Our approach started with collecting relevant papers. As a seed, we collected papers from the recent literature review conducted by Watson ~\cite{watson2020deep}.
    \item \textbf{Search:} To this list, we added papers added by our own searches on Google Scholar. Our search keywords included ``deep learning AND software", ``deep learning AND defect prediction", and ``deep learning AND bug fix" (this last criteria was added since we found that some recent papers, such as  \citet{lee2020continual}, used the term ``bug fix time" rather than ``issue close time").
    \item \textbf{Filter:} Next, we filtered papers using the following criteria: (a) published in top venues as listed in Google Scholar metrics for Software Systems, Artificial Intelligence, and Computational Linguistics; or, released on arXiv in the last 3 years or widely cited ($>$ 100 cites) (b) has at least 10 cites per year, unless it was published in or after 2017 (the last three years). The distribution of papers across different venues is shown in Figure \ref{fig:venues}.
    \item \textbf{Backward Snowballing:} As recommended by \citet{wohlin2014guidelines},  we performed ``snowballing'' on our paper (i.e. we added papers cited by the papers in our list that also satisfy the criteria above).
    Our snowballing stopped when  either (a) the list of papers cited by the current generation is a subset of the papers already in the list, or (b) there were no further papers found.
\end{itemize}

This led to a list of 99 papers, which we summarize in Figure~\ref{fig:heatmap-total}. Some engineering judgement was used in assigning papers to the categories of that figure. For example, a paper on learning a latent embedding of an API ~\cite{nguyen2017exploring} for various purposes, such as discovering analogous APIs among third-parties ~\cite{chen2019mining}, was categorized as ``code comprehension". Similarly, most papers performing some variant of code translation, including API translation as in ~\cite{gu2017deepam}, were categorized into ``language processing"--a bin that contains programming language processing and natural language processing. Tasks that we could not justifiably merge into an existing bin (e.g. on image processing ~\cite{ott2018deep, sun2018neural} were given their own special category.  

Note the numbers on top of the columns of Figure \ref{fig:heatmap-total}:
\begin{itemize}
     \item Sightly more than half (60.1\%) of those papers  compare their results to  non-DL methods. We suggest that number should be higher--it is important to benchmark  new methods against prior state-of-the-art.
    \item Only a minority of papers (39.4\%)  performed any sort of hyper-parameter optimization (HPO), i.e., used methods that tune the various ``hyper-parameters", such as the number of layers of the deep learner, to eke out the best performance of deep learning (39.4\%).
    \item Even fewer papers (18.2\%) applied hyper-parameter optimization in a non-trivial manner;  i.e., not  using deprecated grid search~\cite{bergstra2012random} and using a hold-out set to assess the tuning before going to a separate test set). 
    \item Finally, few papers (10.1\%)  used both  non-trivial hyper-parameter optimization and compared to results to prior non-deep learning work.
    These  ``best of breed''
    papers are listed in Table~\ref{tab:papers}.
\end{itemize}

\begin{table}[h]
    \centering
    \caption{Papers in Column (z) of Figure \ref{fig:heatmap-total}.}
    \label{tab:papers}
    \begin{tabularx}{\linewidth}{Ll}
    \toprule
        \textbf{Paper} & \textbf{Reference} \\
        \midrule
        Suggesting Accurate Method and Class Names & \cite{allamanis2015suggesting} \\
        Automated Vulnerability Detection in Source Code Using Deep Representation Learning & \cite{russell2018automated} \\
        A convolutional attention network for extreme summarization of source code & \cite{allamanis2016convolutional} \\
        Automating intention mining & \cite{huang2018automating} \\
        Sentiment analysis for software engineering: How far can we go? & \cite{lin2018sentiment} \\
        500+ times faster than deep learning: A case study exploring faster methods for text mining stackoverflow & \cite{menzies2018500+} \\
        Automatically learning semantic features for defect prediction & \cite{wang2016automatically} \\
        Deep green: Modelling time-series of software energy consumption & \cite{romansky2017deep} \\
        On the Value of Oversampling for Deep Learning in Software Defect Prediction & \cite{yedida2021value} \\
        \bottomrule
    \end{tabularx}
\end{table}
In summary,
we find that the general
pattern in the literature
is that while there is much new work on deep learning,
there is not so much work
on comparing these new methods to older, simpler approaches.
This is a concern since,
as shown in this paper,
those older simpler methods, being faster, are more amenable to hyper-parameter optimization, and can yield better results when tuned. As we stated above, 40\% of papers do \textit{not} compare against simpler, non-deep learning methods, and only 18\% of papers apply hyper-parameter optimization to their approach, possibly due to the computational infeasible nature of doing so with more complex methods.

\section{Discussion and Conclusion}
\label{sec:discussion}

In this paper, we explored the state of literature applying deep learning techniques to software engineering tasks. 
%Through a thorough analysis of the field, we showed a skew towards more recurrent network models, and some concerning trends surrounding the rigor of evaluating deep learning models. 
We discussed and explored a systemic tendency to choose fundamentally more complex models than needed. We used this, and the study by \citet{galke2021forget} as motivation to apply simpler deep learning models to two software engineering tasks, predicting issue close time, and vulnerability detection. Our model is much simpler than prior state-of-the-art deep learning models and takes significantly less time to run. We argue that these ``old but gold" models are sorely lacking in modern deep learning applied in SE, with researchers preferring to use more sophisticated methods.

As to why it performs so well,
 we hypothesize that the power of SIMPLE came from tuning the hyper-parameters. To test this, we also ran a feedforward architecture without tuning (see FF in Table \ref{tab:bugzilla}). We note a stark difference between the performance of the untuned and tuned versions of this architecture. 
 
From our results, we say that deep learning is a promising method, but should be considered in the context of other techniques. We suggest to the community that before analysts jump to more complex approaches, they try a simpler approach; at the very least, this will form a baseline that can endorse the value of the more complex learner. There is much literature on baselines in SE: for example, in his textbook on empirical methods for AI, \citet{cohen1995empirical} strongly advocates comparing against simpler baselines. In the machine learning community, \citet{holte1993very} uses the ``OneR'' baseline to judge the complexity of upcoming tasks. In the SE community, \citet{whigham2015baseline} recently proposed baseline methods for effort estimation (for other baseline methods, see \citet{mittas2012ranking}). \citet{shepperd2012evaluating} argue convincingly that measurements are best viewed as ratios compared to measurements taken from some minimal baseline system. Work on cross versus within-company cost estimation has also recommended the use of some very simple baseline (they
recommend regression as their default model \cite{kitchenham2006systematic}).

Our results present a cautionary tale about the pitfalls of using deep learners. While it is certainly tempting to use the state-of-the-art results from deep learning literature (which, as prior work has shown, certainly yields good results), we advise the reader to instead attempt the use of simpler models and apply hyper-parameter tuning to achieve better performance, faster.

It is left as future work to explore whether this same principle of using SIMPLE models for other software engineering tasks works equally well. By relying on simple architectures of deep learners, we obtain faster, simpler, and more space-efficient models. This exploration naturally lends itself to the application of modern deep learning theory to further simplify these SIMPLE models. In particular, \citet{han2015deep} explored model compression techniques based on reduced-precision weights, an idea that is gaining increasing attention in the deep learning community (we refer the reader to ~\citet{gupta2015deep} and ~\citet{wang2018training} for details, and ~\citet{tung2018clip} for a parallel implementation of these techniques). Further, knowledge distillation ~\cite{hinton2015distilling}, a method of training student learners (such as decision trees) from a parent deep learning model, has shown great promise, with the student learners outperforming the deep learners they were derived from. This would make it possible to have the accuracy of deep learning with the speed of decision tree learning.

To repeat some comments
from the introduction,
  the experiments of this paper are based on two case studies. Hence, they do not show that {\em all} deep learners can be replaced by faster and simpler methods.
  That said,
  we would say that there is enough evidence here to
  give the software analytics
  reasons to pause, and reflect, on the merits of rushing headlong into new
  things without a careful
  consideration of all
  that has gone before.

\section*{Declarations}

\begin{itemize}
\item \textbf{Funding: } None.

\item \textbf{Conflicts of interest/Competing interests: } None.

\item \textbf{Availability of data and material: } All data used in this manuscript is publicly available at \url{https://github.com/mkris0714/Bug-Related-Activity-Logs}.

\item \textbf{Code availability: } All source code used is available at \url{https://github.com/fastidiouschipmunk/simple}.

\end{itemize}
\balance
\bibliographystyle{spbasic}      % basic style, author-year citations
\bibliography{cite}   % name your BibTeX data base

\end{document}